\documentclass[preprint,aps,prd,floatfix,nofootinbib,11pt]{revtex4-2}
\pdfoutput=1
\usepackage{graphicx}
\usepackage{amsfonts}
\usepackage{nicefrac}
\usepackage{xcolor}
\usepackage{natbib}
\usepackage{braket}

\usepackage{hyperref}

\usepackage{amsmath}
\usepackage{rotating}
\usepackage{amssymb,amsthm}
\usepackage{soul}

\usepackage{xcolor}
\usepackage[normalem]{ulem}

\usepackage{latexsym}
\usepackage{graphics}

\usepackage{amsfonts}
\usepackage{amsmath}
\usepackage{rotating}
\usepackage{amssymb}

\usepackage{amsmath}
\usepackage{rotating}
\usepackage{amssymb}
\usepackage{soul}

\usepackage{xcolor}

\usepackage{xcolor}
\def\beq{\begin{eqnarray}}  
\def\eeq{\end{eqnarray}}

\begin{document}

\title{The quantum Hall effect under the influence of gravity and inertia: A unified approach}



\author{A. Landry}
\email{a.landry@dal.ca}
\affiliation{Department of Mathematics and Statistics, Dalhousie University, Halifax, Nova Scotia, Canada, B3H 3J5}

\author{F. Hammad}
\email{fhammad@ubishops.ca}
\affiliation{Department of Physics and Astronomy, Bishop's University, 2600 College Street, Sherbrooke, QC, J1M\,1Z7 Canada}
\affiliation{Physics Department, Champlain College-Lennoxville, 2580 College Street, Sherbrooke, QC, J1M~2K3 Canada}

\author{R. Saadati}
\email{rsaadati@ubishops.ca}
\affiliation{Department of Physics and Astronomy, Bishop's University, 2600 College Street, Sherbrooke, QC, J1M\,1Z7 Canada}

\begin{abstract}

The quantum Hall effect under the influence of gravity and inertia is studied in a unified way. We make use of an algebraic approach, as opposed to an analytic approach. We examine how both the integer and the fractional quantum Hall effects behave under a combined influence of gravity and inertia using a unified Hamiltonian. For that purpose, we first re-derive, using the purely algebraic method, the energy spectrum of charged particles moving in a plane perpendicular to a constant and uniform magnetic field either (i) under the influence of a nonlinear gravitational potential or (ii) under the influence of a constant rotation. The general Hamiltonian for describing the combined effect of gravity, rotation and inertia on the electrons of a Hall sample is then built and the eigenstates are obtained. The electrons mutual Coulomb interaction that gives rise to the familiar fractional quantum Hall effect is also discussed within a such a combination.

\end{abstract}

\maketitle


\section{Introduction}\label{sec:Intro}
The possible influence of a nonlinear gravitational potential on the integer quantum Hall effect was recently studied in Ref.\,\cite{GQHE}. It was shown there that, unlike the effect of the linear gravitational potential of the Earth \cite{Hehl}, that might be viewed as giving rise only to an apparent modified Hall resistivity, the nonlinear gravitational potential created along the equatorial plane inside a massive solid sphere of uniform mass density affects the degeneracy of Landau levels in such a way that the Hall resistivity is unambiguously affected. This comes about thanks to the simple harmonic gravitational potential inside of which the charge carriers are immersed. In Ref.\,\cite{GQHE}, only the integer quantum Hall effect under the influence of gravity was thus studied. 

However, besides the integer quantum Hall effect, which was the first to be experimentally discovered \cite{PRL1980}, the fractional quantum Hall effect discovered soon after \cite{PRL1982,PRL1987} gave rise to tremendous new theoretical developments soon after its theoretical explanation \cite{Laughlin1983}. See, e.g., Refs.\,\cite{Tong,Yoshioka,ChakrabortyPietiUiinen} for a comprehensive textbook introduction to the subject, see the review \cite{Feldman2021} for the more modern developments, and see Ref.\,\cite{Klitzing2024} for a review of the modern applications and experimental results. 

In the present work, we aim at exploring the fate of both the fractional and integer quantum Hall effects not only in the presence of the nonlinear gravitational potential inside a massive sphere considered in Ref.\,\cite{GQHE}, but also their fate in the presence of inertial effects caused by a uniform rotation of the Hall sample. In fact, inertial effects on the quantum Hall effect have already been examined in Refs.\,\cite{Johnson2000,Fischer2001,Brandao2015,Brandao2017,Lima2018,Matsuo2011,PRD2012}. By solving the corresponding Schr\"odinger equation, it was found that the energy spectrum of the charge carriers of the sample is affected in a nontrivial way. The results from those works rely on the equivalence principle at the quantum level \cite{AC1973,Harris1980}, which finds thus a very interesting application among many others reported in the literature within and outside of condensed mater physics (see e.g., Refs.\,\cite{Tidal,Frontiers2022,Dalabeeh-Sandouqab,Electronics2023,DasPandaDas,Ruggiero2023a,Ruggiero2023b,CanJ2024,Lammerzahl&Ulbricht} and references therein.) 

However, our goal in this paper is not so much to illustrate the equivalence principle at the quantum level as to study the combined effect of gravity and inertia on the motion of charged particles. Furthermore, our investigation here is restricted to Newtonian gravity, as opposed to the studies reported in Refs.\,\cite{Can2014,LandauInSchwarzschild,LandauInKerr,LandauInsideSchwarzschild,2DantiDeSitter,Indian2023} where gravity is mainly taken in its general relativistic expression; i.e., as a manifestation of the curvature of spacetime. 

The remainder of this paper is organized as follows. In Sec.\,\ref{sec:Gravity}, we formulate in terms of creation and annihilation operators the Hamiltonian that describes the dynamics of the electrons of the Hall sample as they move under the influence of the uniform magnetic field and the nonlinear gravitational potential along the equatorial plane of a massive solid sphere. In Sec.\,\ref{sec:Inertia}, we formulate in terms of those operators the Hamiltonian of the electrons under the influence of the uniform magnetic field and pure inertia by letting the Hall sample spin around the axis of the magnetic field. The Hamiltonian of the electrons moving under a simultaneous influence of gravity and inertia is then built in Sec.\,\ref{sec:Gravity-Inertia}. The eigenstates of such a Hamiltonian are given in Sec.\,\ref{sec:Eigenstates}. An outlook on the eventual amplification of the effect due to the induced electric field caused by the compression of the atoms of the sample under gravity and inertia is given in Sec.\,\ref{sec:Outlook}. We conclude this paper with a brief summary and discussion section in which we comment and highlight our main findings.
\section{The Hall sample under gravity}\label{sec:Gravity}
The effect of a nonlinear gravitational potential is achieved by putting the Hall sample between two identical massive neutral hemispheres, each having  the same radius $R$. However, as the thickness of the sample is extremely small compared to the radius of the massive hemispheres, one can, to a very good approximation, consider the electrons of the sample to be effectively moving along the equatorial plane inside the gravitational potential of a full solid sphere. For such a case, the gravitational potential $V_g(x,y)$ at any point of coordinates $(x,y)$ from the center of the sphere can easily be shown to be given by $V_g(x,y)=-\frac{2}{3}\pi m_eG\rho\left(R^2-x^2-y^2\right)$ \cite{GQHE}, where $m_e$ is the mass of the electron, $G$ is Newton's constant and $\rho$ is the uniform mass density of the solid sphere. 

Denoting the charge of the electron by $-e$, the Hamiltonian describing the dynamics of the latter then reads $H=({\bf p}+e{\bf A})^2/2m_e+V_g$. Taking the magnetic field ${\bf B}=B\hat{z}$ to be parallel to the $z$-direction which is chosen to be along the axis of the sphere that is perpendicular to the equatorial plane of the latter, and choosing the symmetric gauge, for which the vector potential has the form ${\bf A}=\frac{1}{2}B\left(-y,x,0\right)$, the Hamiltonian in Cartesian coordinates then reads, up to an unimportant additive constant:
\begin{align}\label{GraHamiltonian}
H&=\frac{1}{2m_e}\left({\bf p}+e{\bf A}\right)^2+m_e\frac{2\pi G\rho}{3}\left(x^2+y^2\right)\nonumber\\   
&=\frac{1}{2m_e}\left(p_x-\tfrac{1}{2}eBy\right)^2+\frac{1}{2m_e}\left(p_y+\tfrac{1}{2}eBx\right)^2+m_e\Lambda_{\rm G}\left(x^2+y^2\right).
\end{align}
We have set $\frac{2}{3}\pi G\rho=\Lambda_{\rm G}$, where the subscript `$\rm G$' denotes here a gravity-induced parameter. In addition, let us introduce the following four ladder operators:
\begin{align}\label{LadderOperators}
    a_L&=\frac{1}{4}\sqrt{\frac{2m_e\varpi_{\rm G}}{\hbar}}\left(x+iy\right)+i\frac{p_x+ip_y}{\sqrt{2\hbar m_e\varpi_{\rm G}}},\qquad a_R=\frac{1}{4}\sqrt{\frac{2m_e\varpi_{\rm G}}{\hbar}}\left(x-iy\right)+i\frac{p_x-ip_y}{\sqrt{2\hbar m_e\varpi_{\rm G}}},\nonumber\\
    a_L^{\dagger}&=\frac{1}{4}\sqrt{\frac{2m_e\varpi_{\rm G}}{\hbar}}\left(x-iy\right)-i\frac{p_x-ip_y}{\sqrt{2\hbar m_e\varpi_{\rm G}}},\qquad a_R^{\dagger}=\frac{1}{4}\sqrt{\frac{2m_e\varpi_{\rm G}}{\hbar}}\left(x+iy\right)-i\frac{p_x+ip_y}{\sqrt{2\hbar m_e\varpi_{\rm G}}},
\end{align}
where 
\begin{equation}\label{InducedFrequency}
\varpi_{\rm G}=\sqrt{\omega_c^2+8\Lambda_{\rm G}}.   
\end{equation} 
For convenience, here and henceforth, we set $eB/m_e=\omega_c$. The non-vanishing commutation relations satisfied by these four operators are: 
$[a_L,a^{\dagger}_L]=1$, $[a_R,a^{\dagger}_R]=1$. The Hamiltonian (\ref{GraHamiltonian}) then takes the following expression in terms of the ladder operators (\ref{LadderOperators}):
\begin{equation}\label{LadderHamiltonian}
H=\tfrac{1}{2}\left(a_R^{\dagger}a_R+a_L^{\dagger}a_L+1\right)\hbar\varpi_{\rm G}+\tfrac{1}{2}\left(a_R^{\dagger}a_R-a_L^{\dagger}a_L\right)\hbar\omega_c.
\end{equation}
The angular frequency $\omega_c$ is the usual one that would be obtained for the electrons under the influence of the magnetic field in the absence of gravity and in a non-rotating sample. The second term in this Hamiltonian is simply $\tfrac{1}{2}\omega_cL_z$, where $L_z=xp_y-yp_x$ is the angular momentum operator. 

Let the eigenvalues of the operators $N_R\equiv a_R^{\dagger}a_R$ and $N_L\equiv a_L^{\dagger}a_L$ be the positive integers $n_R$ and $n_L$, respectively. These two operators can be viewed as number operators for right and left movers, respectively. Then, by setting $n_R+n_L=n$ and $n_R-n_L=\ell$, the energy eigenvalues of the Hamiltonian (\ref{LadderHamiltonian}) read, $E_{n,\ell}=\tfrac{1}{2}(n+1)\hbar\varpi_{\rm G}+\tfrac{1}{2}\ell\hbar\omega_c$. We thus recover the complete splitting of the Landau levels under the influence of the nonlinear gravitational potential, already derived in Ref.\,\cite{GQHE} where only the positive quantum number $\ell$ was considered. Our expression here generalizes the result in Ref.\,\cite{GQHE} as $\ell=n_R-n_L$ can now be positive, negative or zero. Clearly, the degeneracy of the Landau levels is in this case completely lost because of the gravitational field.

Turning off the gravitational field leads to $\varpi_{\rm G}=\omega_c$, for which the $E_{n,\ell}$ take the usual form of the infinitely degenerate quantized Landau levels caused by a pure magnetic field: $E_{n}=\hbar(n_R+\tfrac{1}{2})\omega_c$. On the other hand, turning off the magnetic field leads to $\varpi_{\rm G}=\sqrt{8\Lambda_{\rm G}}$, for which the $E_{n}$ take the usual form of the quantized levels of the two-dimensional simple harmonic oscillator: $E_{n}=(n_R+n_L+1)\hbar\omega_0$, of fundamental frequency $\omega_0=\sqrt{2\Lambda_{\rm G}}$. Unlike the Landau levels, these gravitational levels are, each, $(n_R+n_L+1)$-fold degenerate. For this reason, only extremely high energy levels could come close to mimic Landau levels' infinite degeneracy. Nevertheless, one might still hope to be able to reproduce the famous quantum Hall effect plateaus even in the absence of the magnetic field by simply relying on such a nonlinear gravitational potential.

Before building the eigenstates of the Hamiltonian (\ref{LadderHamiltonian}), we will first derive in the next section the Hamiltonian for the electrons under the influence of inertia and without the gravitational field.

\section{The Hall sample under rotation}\label{sec:Inertia}
The effect of rotation and inertia is easily taken into account by letting the Hall sample undergo a constant spinning of angular velocity $\bf\Omega$ that we assume, for simplicity, to be parallel to the vertical magnetic field's direction, ${\bf B}=B{\hat z}$. 

As shown in Ref.\,\cite{AC1973} (see also Refs.\,\cite{Johnson2000,Fischer2001}), the effect of such a rotation on the electrons of the sample is threefold. First, a Coriolis force, given by $2m_e{\bf v}\times{\bf\Omega}$, is induced. Such a force would emerge from the Hamiltonian if one replaces in the latter the canonical momentum ${\bf p}+e{\bf A}$ of the electron by ${\bf p}+e{\bf A}-m_e{\bf \Omega}\times{\bf r}$. The other force to which the electrons of the sample are subjected due to the rotation of the sample is the centrifugal force, given by $-m_e{\bf\Omega}\times({\bf\Omega}\times{\bf r})$. To take into account this force inside the Hamiltonian, one needs to add in the latter the potential term $-\tfrac{1}{2}m_e({\bf\Omega}\times{\bf r})^2$. The last contribution one needs to account for inside the Hamiltonian, is the induced electric field felt by the electrons within their moving reference frame. This induced electric field appears due to the rotation of the sample inside the uniform magnetic field $\bf B$. The induced electric field in the moving reference frame of the sample is given by $({\bf\Omega}\times{\bf r})\times{\bf B}$, causing an induced electric force on each electron given by $-e({\bf\Omega}\times{\bf r})\times{\bf B}$, which is just the well-known Lorentz force law for moving charges inside a magnetic field. One takes account of this induced electric force inside the Hamiltonian by inserting into the latter the potential term $\frac{1}{2}e\Omega Br^2$. 

Therefore, the combination of the uniform magnetic field and the constant rotation gives rise to the following Hamiltonian:
\begin{align}\label{RotHamiltonian}
H&=\frac{1}{2m_e}\left({\bf p}+e{\bf A}-m_e{\bf \Omega}\times{\bf r}\right)^2-\frac{m_e}{2}\left({\bf \Omega}\times{\bf r}\right)^2+\frac{e}{2}\Omega B r^2\nonumber\\   
&=\frac{1}{2m_e}\left(p_x-\tfrac{1}{2}eB_{\rm I}y\right)^2+\frac{1}{2m_e}\left(p_y+\tfrac{1}{2}eB_{\rm I}x\right)^2+m_e\Lambda_{\rm I}\left(x^2+y^2\right),
\end{align}
where, in the second line, we have adopted again the symmetric gauge for the potential vector $\bf A$ and we have set, for convenience, 
\begin{equation}\label{BILambdaI}
B_{\rm I}=B\left(1-\frac{2\Omega}{\omega_c}\right),\qquad \Lambda_{\rm I}=\frac{\Omega}{2}\left(\omega_c-\Omega\right).
\end{equation}
The subscript `$\rm I$' is used here to denote inertia-related parameters. Note that the Hamiltonian (\ref{RotHamiltonian}) has exactly the same form as the Hamiltonian (\ref{GraHamiltonian}), provided only one substitutes in Eq.\,(\ref{GraHamiltonian}) $B$ with $B_{\rm I}$ and $\Lambda_{\rm G}$ with $\Lambda_{\rm I}$. Therefore, the decomposition of the Hamiltonian (\ref{RotHamiltonian}) in terms of ladder operators also takes exactly the form (\ref{LadderHamiltonian}), provided only one performs the substitutions ($B\rightarrow B_{\rm I}$ and $\Lambda_{\rm G}\rightarrow \Lambda_{\rm I}$) in the expression of the angular frequency $\varpi_{\rm G}$ that becomes $\varpi_{\rm I}$ and in the expression of $\omega_c$ that becomes $\omega_{c\rm I}$ in Eq.\,(\ref{LadderOperators}), such that
\begin{equation}\label{VarpiIomegacI}
    \varpi_{\rm I}=\sqrt{\omega^2_{c_{\rm I}}+8\Lambda_{\rm I}},\qquad \omega_{c_{\rm I}}=\frac{eB_{\rm I}}{m_e}.
\end{equation}

It is worth noting here that by choosing an angular speed of rotation such that $\Omega=\omega_c$, the parameter $\Lambda_{\rm I}$ is made to vanish and $B_{\rm I}$ becomes $-B$, leading to $\varpi_{\rm I}=\omega_{c_{\rm I}}$ (the parameter $B_{\rm I}$ is taken in its absolute value inside $\omega_{c_{\rm I}}$.) This case amounts then to simply inverting the uniform magnetic field in the usual quantum Hall effect. This comes about due to the canceling of the centrifugal force by the induced electric field caused by the noninertial reference frame of the charge carriers. The Coriolis force on the latter is such that the magnetic field's effect is counterbalanced, inducing on the electrons the effect of an inverted magnetic field. The quantum Hall plateaus will, in this case, still be observable. 

On the other hand, choosing an angular speed $\Omega=\frac{1}{2}\omega_c$ leads to $\Lambda_{\rm I}=\frac{1}{2}\Omega^2$ and a vanishing $B_{\rm I}$ and $\omega_{c_{\rm I}}$. This case amounts to exactly balancing the effect of the magnetic field by the Coriolis force on the charge carriers. The effect of the electric force due to the motion-induced electric field and the effect of the centrifugal force combine then into a single effect that is opposite to that of a rotation of angular speed $\frac{1}{\sqrt{2}}\Omega$. As in the case of imposing a nonlinear gravitational potential alone (cf.\,Sec.\,\ref{sec:Gravity}), the quantized levels take here the form $E_{n}=(n_R+n_L+1)\hbar\Omega$, and their degeneracy is lost. The quantum Hall effects are thus destroyed.
\section{The Hall sample under gravity and rotation}\label{sec:Gravity-Inertia}
So far, we have considered separately the effect of gravity and the effect of rotation of the sample on the electrons of the latter as they move inside a uniform and constant magnetic field. However, one can actually combine both the gravitational and the inertial effects. Indeed, one has only to insert the sample inside a solid sphere, of uniform mass density $\rho$, along the equatorial plane of the sphere and let the whole setup spin around the axis of the sphere with a constant velocity $\bf \Omega$ parallel to the axis of the latter. Of course, we refrain from discussing here the serious technological challenges such a setup would obviously present.

For such a combination of gravitational and inertial effects, the resulting Hamiltonian, which we call here the `unified' Hamiltonian, is again given by expression (\ref{LadderHamiltonian}), with the substitutions $B\rightarrow B_{\rm I}$ and $\Lambda\rightarrow \Lambda_{\rm GI}$ inside $\varpi_{\rm G}$ and $\omega_c$, where
\begin{equation}\label{Replacements}
B_{\rm I}=B\left(1-\frac{2\Omega}{\omega_c}\right),\qquad \Lambda_{\rm GI}=\frac{\Omega}{2}\left(\omega_c-\Omega+\frac{2\Lambda_{\rm G}}{\Omega}\right).
\end{equation}
The subscript `$\rm GI$' is used to denote gravity-induced plus inertia-induced parameters. Therefore, the expression of the angular frequencies entering the  decomposition (\ref{LadderHamiltonian}) of the unified Hamiltonian in terms of ladder operators take the following forms, respectively:
\begin{equation}\label{VarpiGIomegacGI}
    \varpi_{\rm GI}=\sqrt{\omega^2_{c_{\rm I}}+8\Lambda_{\rm GI}},\qquad \qquad \omega_{c_{\rm I}}=\frac{eB_{\rm I}}{m_e}.
\end{equation}
Note that by setting $\Omega=0$ in Eq.\,(\ref{Replacements}), we recover the combined effect of the magnetic and gravitational fields, whereas by setting $G=0$, we recover the combined effect of the magnetic and rotation effects. However, this more general combination (\ref{Replacements}) offers yet four other special scenarios. 

The first consists of having $B_{\rm I}=0$ and $\Lambda_{\rm GI}=\frac{1}{2}(\Omega^2+2\Lambda_{\rm G})$, leading to $\omega_{c_{\rm I}}=0$. This is achieved for a rotation of angular speed $\Omega=\frac{1}{2}\omega_c$. The magnetic field is fully balanced in this case by the Coriolis force on the charge carriers. The effects of the electric force, of the centrifugal force and of the gravitational force all combine into a single effect that is equivalent to having a larger ``effective'' gravitational source as it is manifested by the parameter $\Lambda_{\rm GI}$. Because in this case we have $\varpi_{\rm GI}\neq\omega_{c_{\rm I}}$, the degeneracy of the Landau levels is lost, as we shall see in the next section, and the quantum Hall effects are destroyed.

The second scenario consists of choosing $\Omega=\omega_c$, for which case we get back the inverted magnetic field scenario combined with the presence of the gravitational field, such that $\Lambda_{\rm GI}=\Lambda_{\rm G}$. This leads again to the case of the Hall sample under the influence of a gravitational field only without rotation. Because in this case we also have $\varpi_{\rm GI}\neq\omega_{c_{\rm I}}$, the degeneracy of the Landau levels is completely lost too, and the quantum Hall effects are destroyed as well.

The third scenario consists of choosing $\Omega^2=2\Lambda_{\rm G}$, for which case we end up with $\Lambda_{\rm GI}=\frac{1}{2}\Omega\omega_c$. This scenario is achieved thanks to the centrifugal force that exactly balances the gravitational force. Because in this case we have again $\varpi_{\rm GI}\neq\omega_{c_{\rm I}}$, the degeneracy of the Landau levels is completely lost, and the quantum Hall effects are destroyed too.

The fourth scenario is achieved by choosing $\Omega=\frac{1}{2}\omega_c+\frac{1}{2}\sqrt{\omega^2_c+8\Lambda_{\rm G}}$, for which case we recover back the case with a pure magnetic field of ``effective'' algebraic value $B_{\rm eff}=-B\sqrt{1+8\Lambda_{\rm G}/\omega^2_c}$. The ($-$) sign in this expression corresponds to an inverted magnetic field. Among all the scenarios we listed here, this last one is the most interesting. Indeed, in this case we have $\varpi_{\rm GI}=\omega_{c_{\rm I}}$, which means that the infinite degeneracy of the Landau levels is fully restored. Consequently, both the integer and the fractional quantum Hall effects are fully preserved in this scenario as will be shown in detail in the next section.
\section{Eigenstates of the unified Hamiltonian}\label{sec:Eigenstates}
We consider in this section the following unified Hamiltonian, inside of which the substitutions (\ref{Replacements}) should be kept in mind throughout the remainder of this paper to keep working with the most general case of combined gravitational and inertial effects:
\begin{equation}\label{UnifiedHamiltonian}
H=\tfrac{1}{2}\left(a_R^{\dagger}a_R+a_L^{\dagger}a_L+1\right)\hbar\varpi_{\rm GI}+\tfrac{1}{2}\left(a_R^{\dagger}a_R-a_L^{\dagger}a_L\right)\hbar\omega_{c_{\rm I}}.
\end{equation}
The creation and annihilation operators in this expression are given by Eq.\,(\ref{LadderOperators}) with the aforementioned substitutions. The eigenstates $\ket{n_L,n_R}$ of this Hamiltonian are then built by successively acting with the creation operators on the ground state $\ket{0,0}$ as follows: 
\begin{equation}\label{nmStates}
\ket{n_L,n_R}=\frac{a_L^{\dagger n_L}}{\sqrt{n_L!}}\frac{a_R^{\dagger n_R}}{\sqrt{n_R!}}\ket{0,0}.
\end{equation}

In view of our discussion of the fractional quantum Hall effect, however, it is more convenient at this stage to switch to the complex coordinates $z=(x-iy)/\beta$ and $z^*=(x+iy)/\beta$, where $\beta=1/\sqrt{2m_e\varpi_{\rm GI}}$. The ladder operators (\ref{LadderOperators}) then take the following forms:
\begin{align}\label{ComplexLadders}
    a_L&=\frac{1}{4}\left(z^*+8\frac{\partial}{\partial z}\right),\qquad a_R=\frac{1}{4}\left(z+8\frac{\partial}{\partial z^*}\right),\nonumber\\
    a_L^{\dagger}&=\frac{1}{4}\left(z-8\frac{\partial}{\partial z^*}\right),\qquad a_R^{\dagger}=\frac{1}{4}\left(z^*-8\frac{\partial}{\partial z}\right).
\end{align}
From these identities, we easily obtain the ground state $\ket{0,0}$ in the complex coordinates representation by solving both equations, $a_L\ket{0,0}=0$ and $a_R\ket{0,0}=0$. We find the following ground state wavefunction: 
\begin{equation}\label{GroundState}
    \Psi_{0,0}(z,z^*)=\frac{1}{\beta\sqrt{8\pi}}e^{-\frac{|z|^2}{8}}. 
\end{equation}
As the Landau levels' degeneracy is completely suppressed, the lowest Landau level, like all the other levels, contains a single electron state. We can see this by writing the non-degenerate Landau energy levels in the form
\begin{equation}\label{GeneralSpectrum}
    E_{n_R,n_L}=\tfrac{1}{2}\left[n_R\hbar(\varpi_{\rm GI}+\omega_{c_{\rm I}})+n_L\hbar(\varpi_{\rm GI}-\omega_{c_{\rm I}})+\hbar\varpi_{\rm GI}\right].
\end{equation}
The lowest level is found at $n_R=0$ and $n_L=1$. We clearly see from this expression that, apart from setting $\varpi_{\rm GI}=\omega_{c_{\rm I}}$, there is no other way of restoring back to the energy levels their infinite degeneracy. Therefore, both the integer quantum Hall effect and the fractional quantum Hall effect are destroyed. Whenever $\varpi_{\rm GI}$ does not depart much from the value of $\omega_{c_{\rm I}}$, however, as in the case studied in Ref.\,\cite{GQHE} based on a reasonable laboratory-scale massive solid sphere, there is still the possibility of having split energy levels that are so close to each other that they could meaningfully give rise to a survival of the quantum Hall effect thanks to localized states around the sample's defects. Before coming back to such a case, however, we start by considering in what follows the scenario in which the relation $\varpi_{\rm GI}=\omega_{c_{\rm I}}$ holds, a case which allows us to discuss both the integer and the fractional quantum Hall effects.

\subsection{The integer quantum Hall effect}
As discussed in Secs.\,\ref{sec:Inertia} and \ref{sec:Gravity-Inertia}, there are two scenarios  for which the condition $\varpi_{\rm GI}=\omega_{c_{\rm I}}$ could be achieved; the first involves pure inertia, the second involves both inertia and the gravitational field. The first is obtained for a rotation speed $\Omega=\omega_c$ in the absence of the gravitational field, the second is obtained for $\Omega=\frac{1}{2}\omega_c+\frac{1}{2}\sqrt{\omega^2_c+8\Lambda_{\rm G}}$. As for such cases the Landau levels' degeneracy is fully restored, we may simply consider the lowest Landau level as it is customary in the literature on the fractional quantum Hall effect. The infinitely degenerate lowest Landau level is then obtained for $n_R=0$, for which case the corresponding eigenstates of interaction-free electrons read,
\begin{equation}\label{LowestStateIQHE}
    \Psi_{0,n_L}(z,z^*)=\frac{z^{n_{L}}}{\beta\sqrt{ 2^{n_L+3}\pi n_L!}}e^{-\frac{|z|^2}{8}}.
\end{equation}
These wavefunctions, of angular momentum $n_L$, are peaked on a ring of radius $r=\sqrt{2n_L}l_{\rm I}$, where $l^2_{\rm I}=\hbar/eB_{\rm I}$. Therefore, for a circular sample of surface area $S$, the number of available states at the lowest Landau level is roughly $SB_{\rm I}e/h$. Thus, the density of electrons filling $\nu$ Landau levels is given by $n=\nu B_{\rm I}/\Phi_0$, where $\Phi_0$ is the usual quantum of flux $\Phi_0=h/e$. The Hall resistivity $B/ne$ in the presence of gravity and rotation takes then the form
\begin{equation}\label{IGHERhoGI}
    \rho_{xy}=\frac{B}{B_{\rm I}}\frac{h}{\nu e^2}=-\frac{h}{\nu e^2}\left(1+\frac{8\Lambda_G}{\omega_c^2}\right)^{-\frac{1}{2}}.
\end{equation}

We see from this formula that without the gravitational source and for the rotation speed $\Omega=\omega_c$ we get back the usual quantum Hall effect, but with an inverted effective magnetic field. In the presence of the gravitational source, however, the result (\ref{IGHERhoGI}) shows that, in contrast to the pure magnetic-field quantum Hall effect, for $\Omega=\frac{1}{2}\omega_c+\frac{1}{2}\sqrt{\omega^2_c+8\Lambda_{\rm G}}$ the variation of the Hall resistivity does depend on the magnetic field, and it does so in a nonlinear way as the parameter $B_{\rm I}$ in this expression also contains $B$ according to Eq.\,(\ref{Replacements}). This is in agreement with what has been reported in Ref.\,\cite{GQHE} based on the purely analytic method and in the absence of rotation. Experimentally, this would entail that for strong magnetic fields/weak nonlinear gravitational potentials, one does recover the usual behavior of the Hall resistivity as the gravitational field's contribution becomes negligible.
\subsection{The fractional quantum Hall effect}
The fractional quantum Hall effect arises thanks to the Coulomb repulsion force between the free electrons of the Hall sample. Therefore, adding such an interaction to the gravitational, inertial and magnetic forces already included in the Hamiltonian in Sec.\,\ref{sec:Gravity-Inertia} amounts to adding the Coulomb potential terms $e^2/(4\pi\epsilon_0|\bf{r}_i-\bf{r}_j|)$ for any two electrons at positions $\bf{r}_i$ and $\bf{r}_j$. However, when taking into account the multi-electron Coulomb interaction, the Hamiltonian of the degenerate Landau levels in the case $\varpi_{\rm GI}=\omega_{c_{I}}$ allows us to simply make use of Laughlin's wavefunction for the electrons in the lowest Landau level, with an odd integer $q$ and for a total number of electrons $N$:
\begin{equation}\label{LowestStateFQHE}
    \Psi_q(z_1,...,z_N)=\prod_{i<j} (z_i-z_j)^q \exp\left(-\sum\limits_{i=1}^N\frac{|z_i|^2}{8}\right).
\end{equation}
For a given electron at position $z_i$, its wavefunction peaks on a ring of radius $r\sim\sqrt{2qN}l_{\rm I}$, leading, when the lowest level is completely filled, to a total number of available states given roughly by $SB_{\rm I}/\Phi_0\sim qN$. The filling fraction then comes out to be the usual fraction with an odd denominator $\nu=N/qN=1/q$. By the same token, the density of electrons at such a level-filling fraction being $n=\nu B_{\rm I}/\Phi_0$, leads to the Hall resistivity $B/ne$ in the presence of gravity and rotation given by
\begin{equation}\label{FGHERhoGI}
    \rho_{xy}=-\frac{h}{(1/q)e^2}\left(1+\frac{8\Lambda_G}{\omega_c^2}\right)^{-\frac{1}{2}}.
\end{equation}
Although we see from this formula that the fractional Hall resistivity also depends in a nonlinear way on the magnetic field, given that the fractional quantum Hall effect is observed for strong magnetic fields this result also shows that the gravitational source has to have a large mass density for the ratio inside the parentheses in Eq.\,(\ref{FGHERhoGI}) to be meaningful.

Note that although the correction term $8\Lambda_{\rm G}/\omega_c^2$ in expressions (\ref{IGHERhoGI}) and (\ref{FGHERhoGI}) of the Hall resistivity is extremely small, it is not excluded that one can bring the effect into evidence even by using the currently available experimental devices as already pointed out in Ref.\,\cite{GQHE}. To see this, let us assume, as already done in Ref.\,\cite{GQHE}, that the solid sphere to be used is made of a material of mass density that is as large as that of pure platinum ($\rho = 21447$\,kg/m$^3$) or larger. The parameter $\Lambda_{\rm G}$ can then be of the order of $\sim10^{-5}$\,s$^{-2}$. Before we proceed further, however, we should emphasize here that as the gravitational parameter $\Lambda_{\rm G}$ does not depend on the radius of the solid sphere but only on the mass density of the latter, the high costs that would necessarily be associated with the material used to build such a sphere should not be an issue. 

First, going back to the second identity giving $\Lambda_{\rm GI}$ in Eq.\,(\ref{Replacements}), we observe that to be able to keep the two contributions (the gravitational term $2\Lambda_{\rm G}$ and the centrifugal term $-\Omega^2$) at a comparable order of magnitude for such an order of magnitude of $\Lambda_{\rm G}$, one needs to impose rotation speeds of the order of $\sim10^{-2}\,$rad/s. Unfortunately, the Lorentz term in $\Lambda_{\rm GI}$ (the term $\Omega\omega_c$) exceeds those values by many orders of magnitude for the currently applied magnetic fields in the quantum Hall effect. Similarly, the Coriolis term (the correction term $2\Omega/\omega_c$ inside $B_{\rm I}$ in Eq.\,(\ref{Replacements})) is orders of magnitude smaller than unity for the currently applied magnetic fields.   

Next, assume the applied uniform magnetic field to be $B=1\,$T. The electrons' cyclotron frequency is then $\omega_c\sim10^{11}\,$rad/s. The rotation speeds required to achieve the condition $\Omega=\frac{1}{2}\omega_c+\frac{1}{2}\sqrt{\omega^2_c+8\Lambda_{\rm G}}$ for which the degeneracy of the Landau levels is fully restored are then also of the order of $\sim10^{11}\,$rad/s; an order of magnitude which is obviously far from being realistic. Nevertheless, if one were to replace the charge carriers by heavier ions instead of electrons (protons were considered in Ref.\,\cite{GQHE} in the absence of rotation), one can bring down the rotation speeds to the order $\Omega\sim10^3\,$rad/s or smaller for ions that are at least $10^8$ times heavier than electrons. For this case, the correction term  $8\Lambda_{\rm G}/\omega_c^2$ becomes of the order of $\sim10^{-10}$. Notwithstanding these limitations, note that this extremely stringent requirement on the rotation speeds is only necessary for \textit{fully} restoring back the infinite degeneracy of Landau levels and without taking into account the amplification due to the induced electric field. In fact, as we shall see in the Sec.\,\ref{sec:Outlook}, a considerable amplification of the gravitational and inertial effects is obtained when one takes into account the induced electric field coming from the compressed atoms of the lattice. In addition, for small rotation speeds, or in the absence of rotation altogether, the degeneracy is lost due to the nonlinear gravitational potential, but an interesting nontrivial effect can still be witnessed at very low temperatures as we shall discuss in what follows.

\subsection{Destroyed quantum Hall effects}
As we argued below formula (\ref{GeneralSpectrum}), both the integer quantum Hall effect and the fractional quantum Hall effect should in principle be destroyed as long as the condition $\varpi_{\rm GI}=\omega_{c_{\rm I}}$ that guarantees the degeneracy of each Landau level is not satisfied. Nevertheless, whenever $\varpi_{\rm GI}$ does not depart significantly from the value of $\omega_{c_{\rm I}}$, the term $\frac{1}{2}\hbar n_L(\varpi_{\rm GI}-\omega_{c_{\rm I}})$ in the energy spectrum (\ref{GeneralSpectrum}) can be viewed as a correction term that merely induces a ``broadening'' of the energy levels $E_{n_R}=\frac{1}{2}\hbar(n_R+1)\varpi_{\rm GI}$ that can thus be viewed as the ``principle'' Landau levels. 

Indeed, for very low temperatures, such a correction term does not depart much from the usual thermal broadening of the Landau levels, which is of order $\Delta_{\rm th}\sim k_BT$ \cite{GQHE}. The effect of such a broadening has been investigated in detail in Ref.\,\cite{GQHE} and shown to induce a smoothing in the transition between the Hall plateaus. The difference is that now the requirement to achieve a comparable effect to a thermal broadening of the Landau levels is made less stringent thanks to the combined rotational and gravitational effects. In fact, realising $\Lambda_{\rm GI}\sim 0$ can now be achieved simply by choosing $\Omega\approx\frac{1}{2}\omega_c+\frac{1}{2}\sqrt{\omega^2_c+8\Lambda_{\rm G}}$, which is relatively (with the ramifications discussed above) much easier to achieve than adjusting the mass density of the gravitational source alone.   
\section{Induced-field amplification}\label{sec:Outlook}
The study we conducted in the previous sections was entirely based on the forces directly exerted on the electrons due to the magnetic field, the gravitational field and the inertial effects. However, given that the free electrons of the sample are swimming in the sea of positive charge coming from the heavy ions of the lattice, any effect on those ions would translate into a modified background potential for the free electrons. 

It was shown in Ref.\,\cite{GQHE} that the nonlinear radial gravitational potential $V_g(r)$ inside the solid field compresses the atoms of the lattice. As a result, an induced radial electric field, given by $E^i(r)\approx(M/7e)\partial V_g/\partial r$, where $M$ is the mass of the atoms of the lattice, also acts on the electrons besides the gravitational field. As such, and to conserve neutrality, an induced electric force is exerted on the free electrons in the same direction and of the same form as the gravitational force, but of a much stronger magnitude than the latter. The effect of the gravitational field becomes thus indirectly amplified. To take account of such an amplification in the absence of rotation, one needs only to substitute the gravitational contribution $\Lambda_{\rm G}=\frac{2\pi}{3} G\rho$ in the angular frequency (\ref{InducedFrequency}) with the induced-field contribution $\Lambda^i_{\rm G}=\frac{2\pi}{21} MG\rho/m_e$ \cite{GQHE}. We see indeed that, as the ratio $M/m_e$ is of the order of $\sim10^5$ for a sample made of copper of atomic mass $M=1.06\times10^{-25}\,$kg, the gravitational contribution could be amplified by at least five orders of magnitude. 

Similarly, in the case of rotation in the presence of gravity the atoms of the lattice are also subject, besides the gravitational force, to the various induced and noninertial forces which the free electrons experience in their moving reference frame. As a consequence, the ions of the lattice get also compressed by the Coriolis force $2M{\bf v}\times{\bf\Omega}$, by the centrifugal force $-M{\bf\Omega}\times({\bf\Omega}\times{\bf r})$ and by the induced electric force $e({\bf\Omega}\times{\bf r})\times{\bf B}$ caused by the induced electric field felt by the ions within their moving reference frame. We can make use of these various forces acting on the lattice atoms to extract the induced electric field by working out the balance equation as done in Ref.\,\cite{GQHE} for the case of the pure gravitational field. The procedure is as follows. 

First, following Ref.\,\cite{Dessler1968}, we assume the pressure of the electrons of the sample to be $p_e=\frac{2}{3}n\epsilon\propto\frac{2}{3}n^{5/3}$, where the average energy $\epsilon$ of the electron gas of density $n$ (not to be confused with the integer $n$ of the energy levels) is taken to be $\epsilon\propto n^{2/3}$. On the other hand, the gradient $\partial p_e/\partial r$ of the pressure of the electron gas is balanced by the total forces acting on the electron gas and is given by $n_0(eE^i+F_{m_e}^i)$, where $n_0$ is the equilibrium density of the electrons. We denoted here the induced electric field due to the ions by $E^i$, and we denoted the various induced forces acting on the electrons, but that are not due to the ions of the lattice, by $F^i_{m_e}$. Therefore, the balance equation for the electrons reads $\frac{10}{9}\epsilon\partial n/\partial r=-n_0(eE^i+F_{m_e}^i)$. 

Next, applying the same reasoning to the lattice atoms of mass $M$, we conclude that the density of the ions of the lattice under the influence of the induced forces $F_M^i$ obeys the equilibrium equation $C\partial n/(n_0\partial r)=-n_0 F^i_{M}$, where $C$ depends on the sample's elastic properties \cite{Dessler1968}. To preserve neutrality throughout the sample, the two equations for the density $n$ should be identified. This leads then to $E^i=(\gamma F_M^i-F^i_{m_e})/e$, where $\gamma=\frac{10}{9}\epsilon n_0/C$ is a constant that depends on the sample's material. For copper, this constant is $\sim 1/7$. Therefore, by gathering all the forces listed above inside the terms $F_{M}^i$ and $F_{m_e}^i$, the induced electric force acting on the free electrons of the sample reads in vector form:
\begin{equation}\label{Balance}
-e{\bf E}^i=\left[m_e\left(\frac{\gamma M}{m_e}-1\right)\left(2{\bf v}\times{\bf\Omega}-{\bf\Omega}\times\left[{\bf\Omega}\times{\bf r}\right]-\frac{4\pi G\rho}{3}{\bf r}\right)+e\left(\gamma+1\right)({\bf\Omega}\times{\bf r})\times{\bf B}\right].
\end{equation}
We clearly see that the dominant contribution to this induced force for the case of weak magnetic fields comes from those terms that are proportional to $\gamma M/m_e$. Nevertheless, in order to keep our analysis as general as possible, we shall keep all the terms in this expression of the induced force. As a consequence, taking account of the latter in the unified Hamiltonian of the electrons besides the forces we already included in the previous section amounts to replacing the effective parameters (\ref{Replacements}) by the following induced ones:
\begin{equation}\label{InducedReplacements}
B_{\rm I}=B\left(1-\frac{2\gamma M}{m_e}\frac{\Omega}{\omega_c}\right),\qquad \Lambda_{\rm GI}=\frac{1}{2}\Omega\left[\gamma\omega_c-\frac{\gamma M}{m_e}\left(\Omega-\frac{2\Lambda_{\rm G}}{\Omega}\right)\right].    
\end{equation}
These expressions show explicitly the effect of the compression of the atoms of the lattice by the gravitational force and by the frame-dependent forces. This amplification of the effect is very promising, for it provides orders-of-magnitude larger parameters than those obtained without taking the induced electric field into account. 

It is important to note that the amplified parts of the two parameters in Eq.\,(\ref{InducedReplacements}) come exclusively from the those terms arising from inertial effects since the amplification is entirely due to the large difference between the masses of the atoms and of the electrons of the Hall sample. The compression of the ions of the lattice caused by the induced electric field due to rotation is not proportional to the mass of the ions and, hence, has no amplifying effect. On the contrary, such a compression actually reduces the effect of the motion-induced electric field on the electrons as the displacement of the ions creates an opposite electric field.

Note also that all our formulas are given in terms of the electron's vacuum mass $m_e$. However, since the effective electron mass $m_e^*$ inside conductors varies from one material to another, and that it could be made as small as $m_e^*\sim0.01 m_e$ \cite{EffectiveMass}, we immediately see from the expressions of $B_{\rm I}$ and $\Lambda_{\rm GI}$ in Eq.\,(\ref{InducedReplacements}) that the amplifying factor could actually reach an order of magnitude of $\sim10^7$ for such a small effective mass. Therefore, the unrealistic values of the rotation speeds required for restoring the infinite degeneracy of the Landau levels obtained from Eq.\,(\ref{Replacements}) before taking into account the induced electric field, should now be extracted from Eq.\,(\ref{InducedReplacements}). Setting $\Lambda_{\rm GI}=0$ in Eq.\,(\ref{InducedReplacements}), leads to the condition:
\begin{equation}
    \Omega=\frac{m^*_e\omega_c}{2M}\left(1+\sqrt{1+\frac{8M^2}{{m^*_e}^2}\frac{\Lambda_{\rm G}}{\omega_c^2}}\right).
\end{equation} 
This entails that for a cyclotron frequency of the order of $\sim10^{11}\,$rad/s, the rotation speeds to be imposed on the Hall sample are of the order of $\sim10^{4}\,$rad/s. For even weaker magnetic fields, this order of magnitude could be lowered down to the more accessible rotation speeds of $\sim10^{3}\,$rad/s (which are already achievable with biomedical instrumentation and are in wide use in biomedical research \cite{RotationSpeeds}) than the ones inferred from formula (\ref{Replacements}). 
\section{Summary and discussion}\label{sec:Conclusion}
We have examined in this paper the effect of both gravity, rotation and inertia on the integer and the fractional quantum Hall effects using a purely algebraic method. We first built a general Hamiltonian for the electrons of the Hall sample moving inside a uniform magnetic field under the influence of a nonlinear gravitational potential and a constant rotation. We then expressed our Hamiltonian in terms of creation and annihilation operators, and then we extracted and explored the energy spectrum of the electrons for three different scenarios. These are obtained for free electrons that are moving under the influence of the uniform magnetic field combined either with (i) a gravitational field giving rise to a harmonic gravitational potential, (ii) or with a constant rotation of the Hall sample, (iii) or with both a nonlinear gravitational potential and a constant rotation of the sample. Based on the latter more general case, we studied yet different scenarios that arise depending on the value of the rotation speed imposed on the Hall sample. We argued that for an arbitrary rotation speed both the integer and the fractional quantum Hall effects are in general destroyed. This comes about because of the lost degeneracy of the Landau levels. 

We argued that as this lost degeneracy arises because of a correction term that can be very small for certain values of the rotation speed, the quantum Hall effects are affected in the form of a mere broadening of the Landau levels. We showed that for very low temperatures this broadening might even become comparable to the usual thermal broadening of the Landau levels. We saw that achieving this goal is made relatively simpler thanks to the combination of rotation and the gravitation field. 

We subsequently showed that for a very specific value of the rotation speed $\Omega$, given in terms of the value of the imposed uniform magnetic field and on the mass density of the gravitational source, one can fully restore the infinite degeneracy of each Landau level. We found that this restored degeneracy at this specific rotation speed gives rise again to both quantum Hall effects, but that the Hall resistivity departs in both cases from the usual expression it has in the pure quantum Hall effects. We found indeed that the resistivity becomes dependent in a nonlinear way on the imposed magnetic field.

We finally showed that although the combined effects of gravity and inertia derived in Sec.\,\ref{sec:Gravity-Inertia} might be extremely weak experimentally, and requiring unrealistic rotation speeds, taking into account the dynamics of the heavy ions of the lattice induces a realistic amplification of the effect; both with and without rotation. The amplification comes about as a result of the compression of the atoms of the lattice which, in turn, induce an electric field that affects the free electrons more than does the gravitational field or the frame-dependent forces like the Coriolis force and the centrifugal force. We showed that the amplification factor for a Hall sample made of atoms of mass $M$ is of the order of $M/m_e$, where $m_e$ is the mass of the free electrons. We argued that it is the compression caused to the atoms of the lattice by gravity and inertia that is  responsible for the amplification, not the induced electric field in the reference frame of the Hall sample. Our present study brings thus a fresh way of combining inertial effects with gravitational effects on charged quantum particles. The possible novel future uses of the quantum Hall effect based on the results derived here, as well as possible ways to experimentally overcome the weakness of the effects, will be the subject of a future work.   
\vspace{6pt}


\section*{Acknowledgements}

A.\,L. is supported by an Atlantic Association for Research in the Mathematical Sciences (AARMS) Fellowship. F.\,H. and R.\,S. are supported by the Natural Sciences and Engineering Research Council of Canada (NSERC) Discovery Grant No. RGPIN-2017-05388, and by the Fonds de Recherche du Québec - Nature et Technologies (FRQNT).



%


\end{document}